\documentstyle[11pt,newpasp,twoside,epsf]{article}
\def\edcomment#1{\iffalse\marginpar{\raggedright\sl#1\/}\else\relax\fi}
\reversemarginpar

\begin{document}

\title{The Physics of Blazar Optical Emission Regions II: 
Magnetic Field Orientation, Viewing Angle and Beaming
}

\author{Michael J. Yuan$^1$, Hien Tran$^2$, Beverley Wills$^1$, D. Wills$^1$}
\affil{$^1$University of Texas at Austin, $^2$Johns Hopkins University}

\begin{abstract}
For a sample of 51 blazars with extensive optical polarization
data, we used circular statistics to calculate the scatter among
the polarization
position angles for each object. We found that this scatter is correlated
with the radio core dominance.
We compared this relationship with the predictions of 
a simple transverse shock
model. The result suggests that blazar jets are likely to cover a wide range of
speeds consistent with those derived from observations of
superluminal motion.
\end{abstract}

\section{Introduction}

Jet are very important for AGN unification. All radio loud objects have 
powerful jets. In the orientation Unification Scheme, blazars, with 
their extreme flux and polarization variability, have the jets
beamed towards us.  However viewing angle unification is not 
the complete story. Do all the jets have similar physical properties
on average, so that orientation is able to explain most of the variations of
observed properties among blazars? 
%Furthermore, if jets can take an wide range of 
%physical parameters, could they
%form a continuous distribution from the very weak radio quiet ones to the
%powerful blazars?

Optical polarization of blazars provides magnetic field information
very close to the jet formation region and is therefore a very important 
tool for probing jet physics. 
%A powerful way to investigate jet properties is to select a sample
%of complete and unbiased radio loud AGNs and study the physics of those 
%jets statistically. However, such sample is very difficult to get.
%In this paper, we will instead study a sample of blazars, which are the
%objects showing the strongest jet observational features and therefore
%easy to have accurate measurements.
We have investigated a sample of 51 blazars with many optical 
polarization measurements 
made over the past 20-30 years. In paper I, we have investigated the
relationship between the optical polarization position angle
and the VLBI radio structure. The results are consistent with 
a shock model.
In this paper, we focus on the variability of optical polarization
position angles. Assuming a simple shock model, we link 
the angle variability with jet orientation and
speed to explore the jet physical parameters among blazars. 

\section{Data and Analysis}

\subsection{Optical Polarization}

%The optical polarization and VLBI data collection have been described 
%in paper I. We also gathered 5GHz rest frame core-lobe ratios 
%(core dominance R) for each object because it is thought to be a good 
%indicator of orientation and Doppler boosting effects. 
%
%Circular statistics is important 
%when position angles can vary by more than $\pi/2$, which is the
%case for our optical polarization data. Thus we
%use the ``circular standard deviation'' $c(\theta)$ to
%represent the scatter in polarization position angle $\theta_i$ 

Since a blazar's optical polarization swings over a wide range of
angles, there exists a 180\deg\ ambiguity.  Linear statistics
based on normal distributions is not strictly correct.  Therefore we use
circular statistics -- a technique little used in astronomy but commonly used
for example, to analyze directions of returning homing pigeons, or wind
directions.  
Thus, from the polarization position angles $\theta_i$, we calculate the
circular standard deviation $c(\theta)$ about the mean angle, to represent 
their scatter (Fisher 1993). Polarization data are axial, i.e. only cover the
range $0^{\circ}$ to $180^{\circ}$. However, the $2\theta_i$ are truly circular
data. Therefore, we first calculate $c(2\theta)$, then divide by 2 to 
get $c(\theta)$: 
%\begin{displaymath}
%c(\theta) = \left[-2\log\left(\frac{\sqrt{(\sum_{i=1}^{n}\cos\theta_{i})^2 + (\sum_{i=1}^{n}\sin\theta_{i})^2}}{n}\right)\right]^{\frac{1}{2}}  
%\end{displaymath}
\begin{displaymath}
c(\theta) = \left[-\ln\left(\frac{\sqrt{(\sum_{i=1}^{n}\cos2\theta_{i})^2 + (\sum_{i=1}^{n}\sin2\theta_{i})^2}}{n}\right)\right]^{\frac{1}{2}}  
\end{displaymath}
where $n$ is the total number of measurements. 
If the sample has a large intrinsic scatter, the determination of
the real scatter depends on the accurate shape of the wrapped distribution
between $-\pi/2$ and $\pi/2$. If the sampling of data is not enough to
reveal the detailed shape of the distribution, the $c(\theta)$ calculation
algorithm gives a maximum $c(\theta) \sim 66^{\circ}$.

\subsection{Radio Core Dominance}

%In the unification of flat-spectrum core-dominated
%and the steep-spectrum lobe-dominated radio sources, Doppler
%boosting of the core synchrotron radiation plays an very important role
%(Blandford \& Rees 1978).
%The relativistic beaming is supported by detection of
%superluminal motions in many bright core-dominant sources.
%Statistical analyses of radio source samples also supported the hypothesis
%of beamed radiation in the bright cores (Orr \& Browne 1982).

In the unification scheme of powerful radio sources, the core-dominated 
sources are the Doppler-boosted jet, viewed end-on. At larger angles
the core becomes much weaker and collimated jets link the well resolved
double lobes. Doppler boosting in core-dominant sources is supported by their
very high brightness temperatures and by superluminal motions. Statistical
analyses of radio source samples support the unified scheme (M. Chiaberge,
this conference).
The ratio of core-to-lobe luminosity, or the core dominance, R, 
is therefore a good indicator of viewing angle. However,
in blazar samples, viewing angle is not the only parameter affecting
Doppler boosting.
From work of other authors (Orr and Browne 1982),
we know that R depends on three physical parameters and can be
expressed as
$$\log\mbox{R} = \log(\mbox{R}_T \times (1-\beta \times \cos(\phi))^{-3})$$
where $\phi$ is the viewing angle of the jet, $\beta$ is the bulk speed of the
emitting electrons in the jet and
$R_T$ is the tangential value of R measured for
an average edge-on jet in radio galaxies.  
We used $R_T=0.0024$ at 5GHz rest frequency
(e.g. Figure~1 of Hoekstra, Barthel \& Hes 1997).
While R is of course sensitive to $\phi$, 
for smaller $\phi$ (e.g. $<20^{\circ}$), it
is increasingly sensitive to $\beta$.

For each object we calculated R at 5 GHz rest frequency, using core and lobe
flux densities from literature.
In Section~4, we will calculate log\,R from models
using the formula above and compare the models with observed values.

%% Figure~1 plots the relation of the actual viewing angle $\phi$ with
%% log\,R at a range of reasonable jet speed $\beta$.
%% From the top to the bottom, the lines are for
%% $\beta=0.995,0.99,0.98,0.97,0.96,0.95,0.93$ respectively.
%% The range of $\beta$ here is the range of reasonable jet bulk motion speeds
%% derived from superluminal motion observations (...).
%% The relation between $\phi$ and log\,R is
%% not very dependent on $\beta$ at low log\,R (big $\phi$) range.
%% The correlation is very good for a given beta.
%% But at high log\,R (small $\phi$)
%% range, the functions are sensitive to $\beta$. We will discuss this
%% effect in later sections.

%% \begin{figure}[hp]
%% \plotone{plot/LogR_Phi_model.ps}
%% \caption{       \label{LogR_Phi_model.ps}
%% The model for the radio core dominance log\,R and jet viewing angle
%% relations. X-axis is the viewing angle in degrees and Y-axis is log\,R.
%% Different curves are for different jet speed $\beta$. From the top to
%% the bottom are $\beta=0.995,0.99,0.98,0.97,0.96,0.95,0.93$ respectively. }
%% \end{figure}

\section{Correlation}

Figure~1~(a) shows that there is a correlation between 
$c(\theta)$ and log\,R. The
correlation is significant at the 99.9\% level. 
However, we note that the correlation is not linear. 
For the log\,R$<$2 region, which includes most blazars, 
$c(\theta)$ increases with log\,R, but for
log\,R$>$2, there is a lot of scatter.
No object has $c(\theta)$ larger than 66\deg, as expected
from the algorithms for large scatter and small sample size.

\begin{figure}
\plottwo{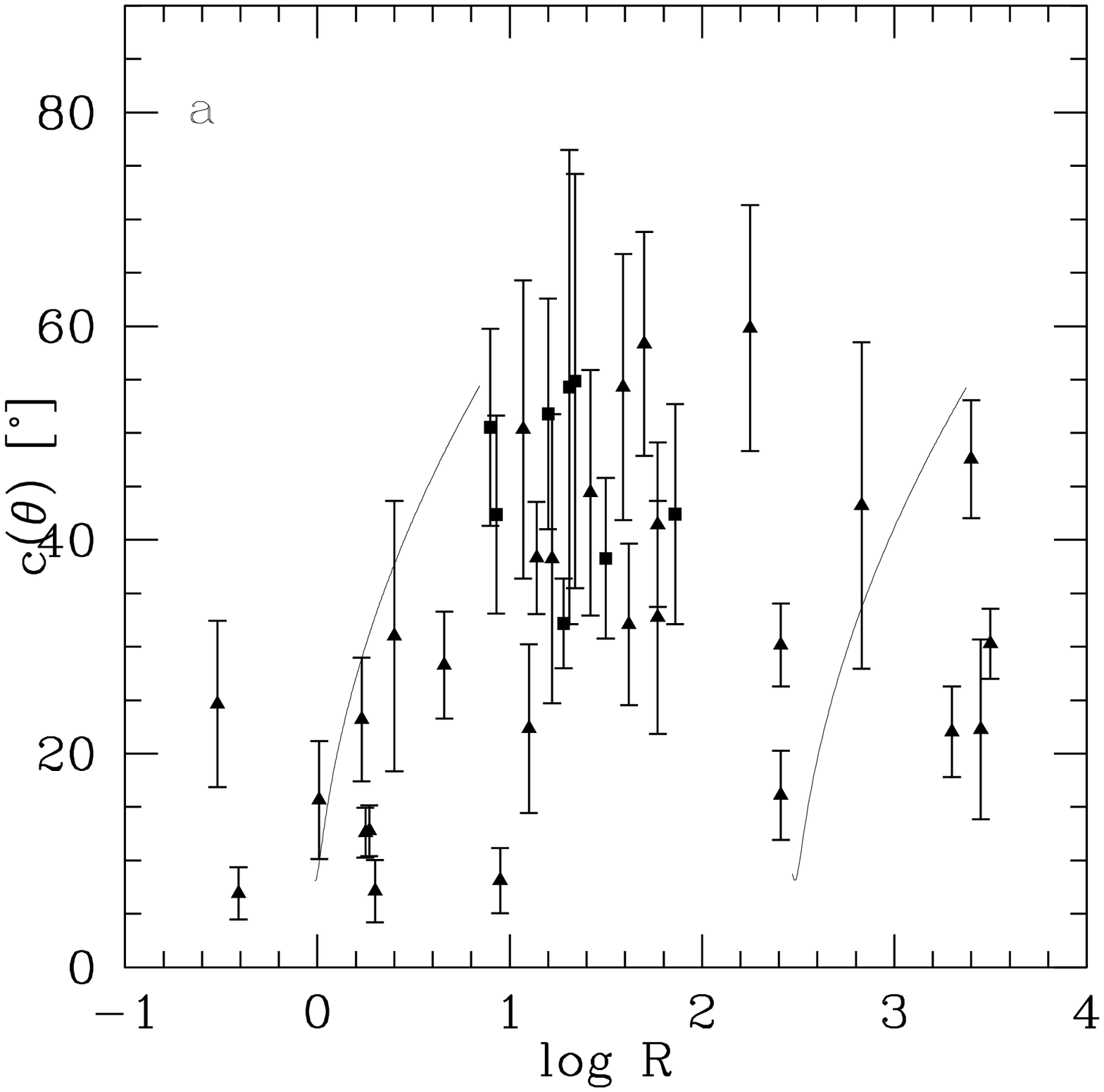}{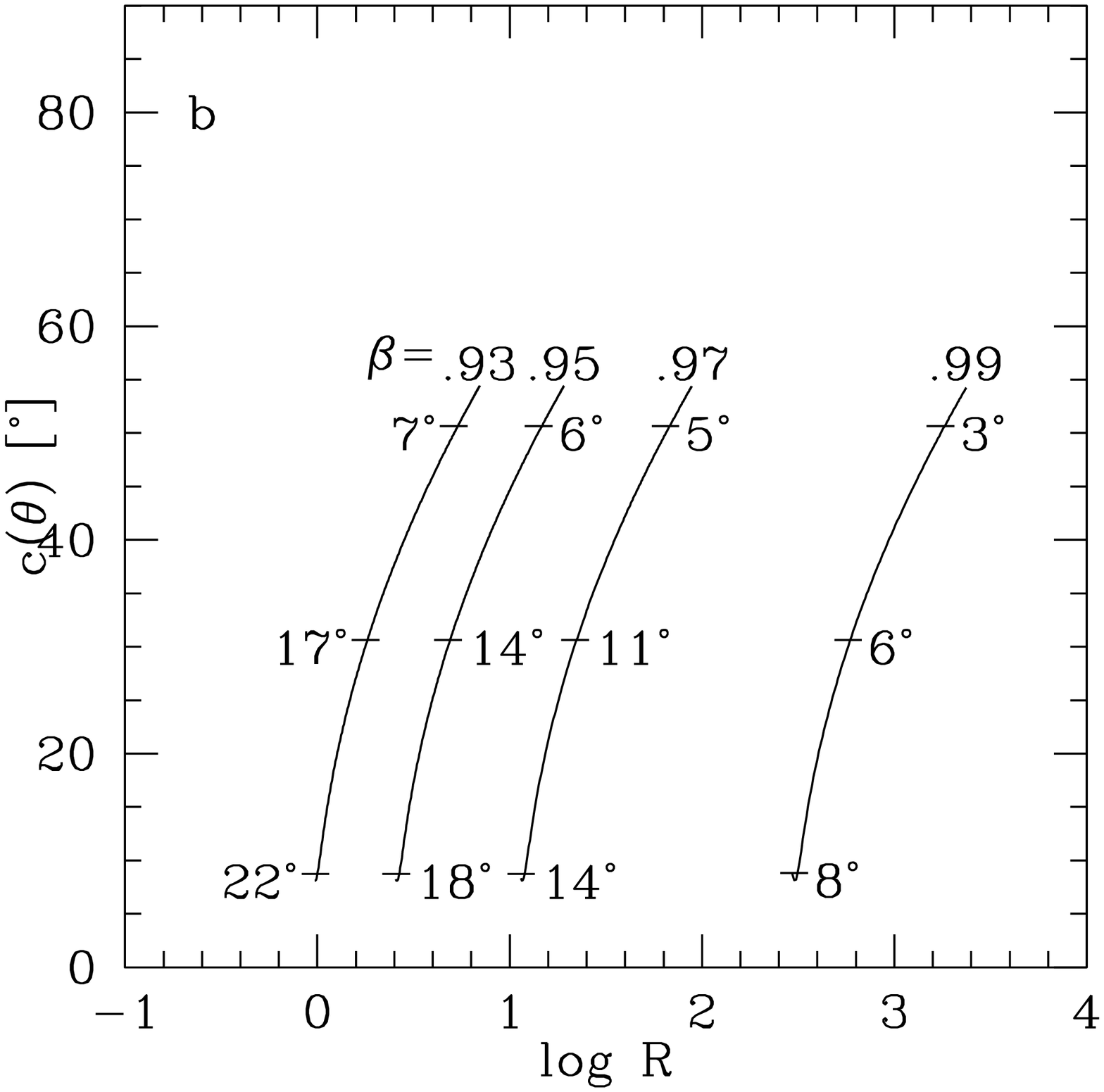}
\caption{(a) -- Correlation between log~R and the scatter of
optical polarization position angle. Triangles are BL Lacs and squares
are QSOs. The
two thin curves represent $\beta=0.93, 0.99$ models. (b) -- Model simulations. 
Each represents represents a model with fixed $\beta$ and variable 
$\phi$. The $\beta$ for each model is noted on top of each curve. The
markers along each curve denote the corresponding $\phi$ in
the observer's frame. }
\end{figure}

\newpage

\section{Model}

The model we use makes the following assumptions:

\begin{enumerate}
\item Optical emission comes from a transverse shock propagating along the 
jet.
\item The shock compresses the magnetic field 
so much that the magnetic component along the jet direction is 
negligible.
\item In the plane of the shock, there is a dominant magnetic field direction 
that changes randomly with time.  %At any given time 
This magnetic field direction corresponds to the emission region 
whose radiation happens to be boosted towards us.  
Different ejection directions observed for emerging blobs on VLBI scales and
complex VLBI structure suggests that these dominant regions change rapidly. 
\item The shock is optically thin.
\end{enumerate}

When the shock is viewed face on, the magnetic field direction projected on the
sky will vary randomly with time.   As the viewing angle is increased, the
projected magnetic field direction will lie increasingly perpendicular to the
projected jet direction.

%Although the magnetic field direction is random in the shock plane, the magnetic
%field direction projected on the sky will appear increasingly
%component parallel to the shock plane 
%from an angle, the direction of the 
%perpendicular to our line of sight. 
%% (Figure 3) 

%We calculate the projected magnetic field directions in the co-moving frame.
A simplification is afforded by the Lorentz invariance of the ratios of Stokes
parameters, so our calculations of projected magnetic field direction 
in the co-moving frame (viewing angle $\phi'$) can be referred to the
observer's frame using the relation (Bjornsson, C.-I. 1982):

$$\sin(\phi') = \frac{\sin(\phi)}{1-\beta\cos(\phi)}\cdot\sqrt{1-\beta^2} $$
%The appearance of the shock is
%further complicated by Doppler boosting.
%However, for polarization, it can be relatively simple.
%The ratios of Stokes parameters are Lorentz invariants.
%The only effect we need to consider to convert the polarization
%observed in co-moving frame to the observer's frame is the aberration.
%We will first convert the viewing angle to co-moving frame using the
%aberration formula.
%Then we take advantage of the simplified geometry in co-moving frame
%and the study polarization angle distributions. The results apply to
%observer's frame directly since the polarization angle
%does not change in Lorentz transformation as we mentioned above.
%
%
%% \begin{figure}
%% \plotone{plot/shock_geometry.ps}
%% \caption{The geometry of the shock magnetic field. Viewed from face on
%% or from an angle $\phi$ in the co-moving frame.}
%% \end{figure}
%For a given viewing angle ($\phi<\sqrt{1-\beta^2}$)
%and speed for the shock, we calculate
%the viewing angle in the shock co-moving frame.
%
%$$\sin(\phi') = \frac{\sin(\phi)}{1-\beta\cos(\phi)}\cdot\sqrt(1-\beta^2) $$.

For a given set of $\beta$ and
$\phi$, we generate 100 randomly oriented magnetic
field directions to simulate the random changes over time inside
the shock plane and project them onto the sky in the co-moving frame.
The circular standard deviation of those angles can then be compared with real 
observations, using the relation between R and $\phi$ given earlier.
$\mbox{R}_T$ is fixed by observations, 
so the only free parameter in this model is
$\beta$.

%in the observer's frame), 
%and compared with real observations. As we have noted
%before, jet speed together with viewing angle is sufficient to determine
%the core dominance log~R.
%
%So, for each set of jet speed and orientation, we can calculate the
%log~R and simulate a circular standard deviation for optical 
%polarization angles.
%The only free parameter in the model is the jet speed.

The results are shown in Figure~1~(b).
Each curve represents a model with fixed $\beta$, with $\phi$ 
decreasing from bottom left to top right. The values of $\phi$ are marked
along the curve.
From left to right, $\beta=0.93, 0.95, 0.97, 0.99$.
The upper limit on $c(\theta)$ is the result of the computational limit.
The lower limits are caused by the fact that we only have
100 simulated vectors to do statistics and any one vector that happens to
be parallel to the jet direction contributes a lot to the scatter. 
In principle, if
we had an increasing number of vectors, the lower limit for $c(\theta)$
would approach 0.
%The lower limit is the result of the limit on viewing angles for a given
%jet speed ($\phi<\sqrt{1-\beta^2}$).
%The observed points lie between $\beta=0.93$ and $\beta>0.990$. 
On our model,
the $c(\theta)$ vs. log\,R correlation arises because $\beta \sim 0.95$ for
most blazars, and so the viewing angle dependence dominates for log\,R$< 2$.
For log\,R$> 2$, the scatter is the result of a tail to higher jet speeds.

%there is a lower limit on jet speed between $\beta=0.93$ to $\beta=0.95$.
%The very strong correlation we observed for log\,R$<$2 region fits the lower
%jet speed models very well while the high jet speed models account for the
%scatter in high log\,R region.  That suggests a rather 
%wide range of jet speed.

Our simple model with only one free parameter, $\beta$, accounts for the 
correlation we find between the scatter in optical polarization angles,
c($\theta$),
and the core-dominance, log~R. 
Figure~1 shows that the average jet speed differs significantly
amongst the powerful blazars,
with a range of values comparable with those derived from superluminal motion
and from the statistics of core-dominance in radio-source surveys (Urry
\& Padovani 1995).
In Unified Schemes it is important to take into account that log R is not
simply a measure of orientation.  Jet speed introduces considerable
scatter, especially at log R $>$ 2.

%Even among blazars showing 
%powerful jets, the jet speeds vary a lot and that contributes significantly
%to the observed optical polarization properties.  Orientation does not 
%count all the differences among blazars.

\end{document}